\documentclass[
aps,
prl,
groupedaddress,
superscriptaddress,
floatfix,
twocolumn,
notitlepage
]{revtex4-1}

\usepackage{graphicx}
\usepackage{amsmath,amssymb}
\usepackage{braket}
\usepackage{hyperref} 
\usepackage{subfigure}
\usepackage{bm} 
\usepackage{mathtools} 
\usepackage{float}

\newif\iffigures

\iffigures
\usepackage{cprotect}
\fi

\begin{document}
	

\title{Comment on ``Gravitational Mass Carried by Sound Waves" [A. Esposito, R. Krichevsky, and A. Nicolis, Phys. Rev. Lett.~\textbf{122}, 084501 (2019)]}
	
\author{D. R. Gulevich} 
\affiliation{ITMO University, St. Petersburg 197101, Russia}
	
\author{F.\,V.~Kusmartsev}
\affiliation{Micro/Nano Fabrication Laboratory Microsystem and THz Research Center, Chengdu, Sichuan 610200, China}
\affiliation{Department of Physics, Loughborough University, Loughborough LE11 3TU, United Kingdom}
\affiliation{ITMO University, St. Petersburg 197101, Russia}

\date{\today}

\begin{abstract} 

\end{abstract}

{\let\newpage\relax\maketitle}


In Ref.~\cite{Esposito} Esposito et al. made an intriguing claim that sound waves carry a nonzero negative gravitational mass -- the effect which suggests consequences for neutron stars, seismic phenomena and even proposed to be detected in the laboratory. The present comment aims to avoid the arising confusion in the scientific community and beyond on how one should interpret Esposito's result. Here, we also provide an additional insight by introducing topological aspects of Esposito's nonlinear excitations which enables us to make important conclusions on conditions necessary for the mass-carrying excitations to be observed. 

\begin{figure}[t!]
	\begin{center}
\includegraphics[width=3.5in]{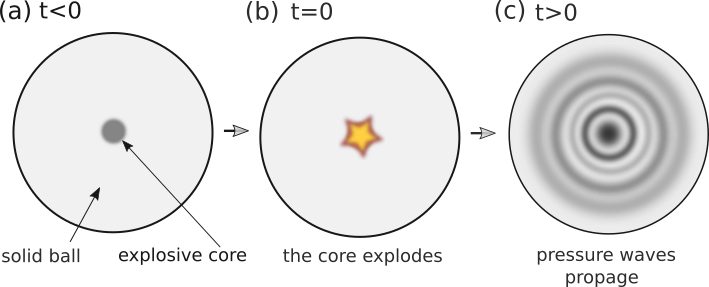}
		\caption{\label{fig:sketch} 
(a) Spherically symmetric solid ball with explosives at the core. (b) At some moment the core explodes and induces (c) At a later time, spherical pressure wave propagates towards the surface and causes the ball to oscillate.
		}		
	\end{center}	
\end{figure}

We will first discuss how the result of Esposito et al. should {\it not} be interpreted. 
The gravitational mass can not be assigned to acoustic waves in the literal sense, that is, as if they were sources of gravitational field. To illustrate this argument, consider a spherically symmetric solid ball (Fig.~\ref{fig:sketch}a). Our following arguments develop in analogy to the Tolman paradox in general relativity~\cite{Tolman,Ehlers-PRD}.
In the center of the ball there is an explosive core. At some moment (Fig.~\ref{fig:sketch}b) the core explodes and produces a spherical pressure waves propagating towards the surface (Fig.~\ref{fig:sketch}c). 
The question arises, whether an external observer can detect the gravitational waves or any change of the gravitational field caused by the generated pressure waves.
If we were to take the claim of Ref.~\cite{Esposito} literally, we should assign the negative gravitational mass to the pressure waves which will lead to a decrease of the gravitational pull at the location of the external observer. This naive suggestion, however, comes in violation to the Birkhoff theorem~\cite{Birkhoff} which states that irrespective of which spherically symmetric changes in a closed spherically symmetric system occur, the metric outside the system will remain the static Schwarzschild metric. Hence, acoustic waves can not be assigned masses in the usual sense, to avoid the conflict with general relativity.




Then, what does the result of Esposito et al. really mean? 
The mass given by the Esposito formula (1) should be interpreted as {\it topological charge} of a 
nonlinear sound excitation 
propagating in solid. To clarify this statement, it is instructive to return to the formula (18) of Ref.~\cite{Esposito} preceding the derivation of the final formula (24) for the negative mass. Based on the Eq. (18), we introduce the topological charge $Q$ by
\begin{equation}
Q = \int q \; d^3\mathbf{r},\quad \text{with}\quad
q=w_0 b_0 \langle \vec\nabla \cdot \vec\pi \rangle.
\label{Q}
\end{equation}
The topological charge~\eqref{Q} defined as an integral of topological charge density $q$ involving first derivatives of the field is ubiquitous in physical systems. Examples include sine-Gordon solitons~\cite{SG-solitons}, magnetic skyrmions~\cite{skyrmions}, exciton-polariton condensates~\cite{sp-Meissner}, topological insulators~\cite{top-in}
where the topological charges can take both integer and continuous range of values.

The significance of the topological charge~\eqref{Q} is best illustrated on a quasi-1D example when the relevant dynamics occurs along a rod of given cross-sectional area $A$. In this case, the expression~\eqref{Q} is easily integrated and yields
$$
Q = A w_0 b_0\left[\pi(\infty) - \pi(-\infty)\right],
$$
which is nothing but the relative dislocation of the material from either side of the wavepacket. Hence, Esposito et al. describe propagation of nonlinear excitations carrying a matter dislocation.
In this respect, Esposito excitations are matter analogues of {\it grey solitons} arising in nonlinear optical media~\cite{Gulevich-SciRep,Ablowitz}: these are characterized by a continuous topological charge and carry a localized density deficiency.



Introducing the topological charge~\eqref{Q} allows us to make an important conclusion: because $Q$ is conserved and is negative, Esposito's grey solitons
can not be excited in the bulk -- neither alone no in pairs, but can only enter via the boundaries~\cite{footnote1}. The practical use of the Esposito formula (24) is to suggest how one can excite such nonlinear excitations: 
to create an excitation of energy $E$ one needs to displace the material to induce an increase in volume by exactly $Q/\rho_m$, where $\rho_m$ is mass density of the medium.

The fact that the Esposito effect occurs in the nonlinear regime is not a coincidence. The displacement of mass is natural to nonlinear phenomena.
Nonlinear excitations such as dark, grey and bright solitons are often associated with the transfer of matter of both positive and negative amounts: some of the known examples of the latter include Langmuir solitons in plasma~\cite{Kusmartsev} and dark solitons on the surface of water~\cite{Chabchoub}.
Esposito result adds one more example to the variety of intriguing nontrivial phenomena arising in nonlinear media.


\begin{thebibliography}{99}

\bibitem{Esposito} 
A. Esposito, R. Krichevsky, and A. Nicolis, 
Phys. Rev. Lett.~\textbf{122}, 084501 (2019).

\bibitem{Tolman}
R. C. Tolman, Relativity, Thermodynamics and Cosmology (Clarendon Press, Oxford, 1934).

\bibitem{Ehlers-PRD}
J. Ehlers, I. Ozsv\'{a}th, E. L. Sch\"{u}cking, and Y. Shang,
Phys. Rev. D~\textbf{72}, 124003 (2005).


\bibitem{Birkhoff}
G. D. Birkhoff, Relativity and Modern Physics (Harvard University, Cambridge, MA, 1923).

\bibitem{SG-solitons}
R. Rajaraman, Solitons and Instantons: An Introduction to Solitons and Instantons in Quantum Field Theory (North Holland, Amsterdam, 1987).

\bibitem{skyrmions}
A. Fert, N. Reyren, and V. Cros,
Nat. Rev. Mater.~\textbf{2}, 17031 (2017).

\bibitem{sp-Meissner}
D. R. Gulevich, D. V. Skryabin, A. P. Alodjants, and I. A. Shelykh,
Phys. Rev. B~\textbf{94}, 115407 (2016).

\bibitem{top-in}
M. Z. Hasan and C. L. Kane,
Rev. Mod. Phys.~\textbf{82}, 3045 (2010).

\bibitem{Gulevich-SciRep}
D. R. Gulevich, D. Yudin, D. V. Skryabin, I. V. Iorsh, and I. A. Shelykh, 
Sci. Rep.~\textbf{7}, 1780 (2017).

\bibitem{Ablowitz}
M. J. Ablowitz, Nonlinear Dispersive Waves: Asymptotic Analysis and Solitons (Cambridge University Press, 2011).

\bibitem{footnote1}
Here we assume an ideal media ignoring any variations of temperature and possible phase transitions.


\bibitem{Kusmartsev}
F. V.  Kusmartsev, Phys. Rep.~\textbf{183}, 1 (1989).


\bibitem{Chabchoub}
A. Chabchoub, O. Kimmoun, H. Branger, N. Hoffmann, D. Proment, M. Onorato, and N. Akhmediev,
Phys. Rev. Lett.~\textbf{110}, 124101 (2013).

\end{thebibliography}
\end{document}